# On the Spatiotemporal Burstiness of Terms

Theodoros Lappas,    Marcos R. Vieira,    Dimitrios Gunopulos,    Vassilis J. Tsotras

University of California, Riverside    University of Athens

{tlappas,mvieira,tsotras}@cs.ucr.edu    dg@di.uoa.gr

## ABSTRACT

Thousands of documents are made available to the users via the web on a daily basis. One of the most extensively studied problems in the context of such document streams is *burst identification*. Given a term $t$, a burst is generally exhibited when an unusually high frequency is observed for $t$. While spatial and temporal burstiness have been studied individually in the past, our work is the first to simultaneously track and measure *spatiotemporal term burstiness*. In addition, we use the mined burstiness information toward an efficient document-search engine: given a user's query of terms, our engine returns a ranked list of documents discussing influential events with a strong spatiotemporal impact. We demonstrate the efficiency of our methods with an extensive experimental evaluation on real and synthetic datasets.

## 1. INTRODUCTION

The World Wide Web serves as a host to overwhelming volumes of documents, appearing in bulk online on a daily basis. Online magazines and newspapers (e.g. nytimes.com), blogging and microblogging platforms (e.g. BlogSpot.com and Twitter.com) and social networking platforms (e.g. Facebook.com) are examples of online venues where users flock to access such documents. In the context of such document streams, one of the most well-studied problems is the identification of bursts. Given a term $t$, a burst is generally identified when an *unusually* high frequency is observed for $t$ in the considered documents. A significant amount of work has been devoted to identifying *temporal* bursts [13, 14, 31, 35]. A temporal burst is typically identified by: **(a)** an interval on the timeline, indicating the specific timeframe during which the unusually high frequency was observed, and **(b)** a score that indicates the burst's strength, i.e. the extent of the deviation from the term's usual frequency. The work on temporal burstiness assumes a single stream of documents. In the context of the web, however, documents are typically associated with a geostamp. In social networking platforms and blogging websites, registered users include their geographical location (i.e. place of residence) as part of their online profile. Further, in news portals such as Topix.com articles are organized based on their place of origin. This setting motivates the study of burstiness in the spatial domain by introducing multiple document streams from different locations. An example of this setting is shown in Figure 1, where the white dots on the map represent different document streams. In recent work, Mathioudakis et al. [18] presented a framework for the identification of *spatial* bursts, where the temporal interval of interest $I$ is given as part of the input (this is a limitation that we overcome in our work). Given such an interval $I$ and a term $t$, the authors focus on identifying geographical regions where the observed frequency of $t$ was unusually high, within the timeframe defined by $I$.

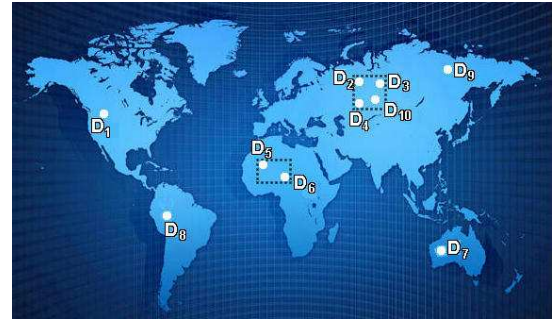

**Figure 1: Spatiotemporal collection $\mathcal{D}$. The white dots represent streams of data originating in different locations on a 2D map.**

In this paper, we present the first framework for simultaneously tracking the spatial and temporal burstiness of terms. In particular, given a term $t$ and a set of document streams from different locations, we focus on two different types of *spatiotemporal burstiness patterns*:

**Combinatorial Patterns:** these patterns ignore the geographical proximity among streams. Instead, they are defined as a combination of a temporal interval and a set of streams, where each stream originates from a different geographical location. Any arbitrary subset of the streams in Figure 1 can be included in a combinatorial pattern (e.g. $\{D_1, D_4, D_7\}$). A combinatorial pattern encodes that *unusually high frequencies were simultaneously observed for term $t$ in all the streams in some set $\mathcal{C}$, during the same temporal interval $I$.*

**Regional Patterns:** these patterns consider the geographical proximity among document streams. They are defined as a combination of a temporal interval and a geographical region. A region can contain the geostamps (locations) of multiple document streams. Two such regions are marked in Figure 1. The 1st region contains streams $\{D_5, D_6\}$, while the 2nd region contains streams $\{D_2, D_3, D_4, D_{10}\}$. This pattern encodes that *unusually high frequencies were observed for term t in geographical region R during a temporal interval I*. In this work, we formalize both of these spatiotemporal patterns and present efficient algorithmic techniques for their identification and evaluation.

**Utilizing spatiotemporal burstiness:** The second part of our work focuses on the utilization of the mined spatiotemporal patterns. In previous work [14], we showed how temporal bursts can be used to identify documents on influential events. In this paper, we present a search engine that considers the spatiotemporal burstiness of terms in the process of document retrieval. Given a query of terms submitted by the user, our engine retrieves *relevant* documents that discuss *events with a major spatiotemporal impact*, i.e. an impact that was reflected in multiple streams for an extended timeframe. We observe that regional patterns lead to documents on events with a strong localized impact. (e.g. a medium-scale earthquake affecting a specific region of the world). One the other hand, combinatorial patterns favor events with a more global effect (e.g. a large-scale pandemic affecting countries across the globe). We demonstrate and discuss these findings in our experiments.

### 1.1 Applications

Next, we list some of the applications of our work:

**Document Search:** Given any query of terms, our search engine can be used to retrieve documents (e.g. tweets on Twitter, news articles, blog-posts) that discuss events with a strong spatiotemporal impact. These are influential events that affect multiple places in the world for extended timeframes. Thus, they are more likely to be of interest to users. Our work is a natural extension of approaches that are limited to the temporal dimension of burstiness [7, 10, 14, 2].

**Document Selection:** by tracking the popularity of terms across space and time, a news portal can determine which articles to present to each user, based on their respective vocabulary, location and timeframe. For example, a user living in region that is being affected by a major event is more likely to be interested in reading relevant articles than someone who lives thousands of miles away.

**Trend Identification:** Spatiotemporal patterns are a natural way to detect trends. Given the set of terms that describe an item (e.g. a person or product), we can identify *when* and *where* it was popular. This information can help the user who is interested in trendy topics and can also be used as input to marketing and advertising campaigns.

### 1.2 Roadmap

The rest of the paper is organized as follows. In Section 2 we introduce some preliminary points. In Sections 3 and 4, we describe the two alternative approaches for identifying spatiotemporal-burstiness patterns. In Section 5, we discuss how the extracted patterns can be used for document search. The experimental evaluation is presented in Section 6. In Section 7 we review the related work. Finally, we conclude the paper in Section 8.

## 2. PRELIMINARIES

In this section we introduce some notation and briefly discuss some basic points that are relevant to our methodology.

**Geostamps and document Streams:** We assume an underlying geographical map and a set of document streams $\mathcal{D} = \{D_1[\cdot], ..., D_n[\cdot]\}$. Here, $D_x[i]$ represents the set of documents reported from stream $D_x$ at timestamp $i$. Each stream is associated with a fixed geographical location (geostamp). For the sake of simplicity (and without loss of generality), our analysis assumes a single streaming source per location (e.g. the aggregated content of all the available blogs or websites in a city).

**Granularity:** Our approaches place no restrictions on the possible locations of the document streams. However, if the number of the considered streams is overwhelming, it can potentially hurt performance. This issue can emerge, for example, when millions of individual users (e.g. on Twitter) are considered as individual streams. For most real-life applications, it is sufficient to consider a stream as an entire city or, at most, a specific neighborhood. Then, users can be easily grouped to form the corresponding aggregate streams. Still, if one chooses a finer granularity, it is preferable to define the problem in the context of the region of interest, instead of considering the entire map. An alternative way to group users is by using a grid to partition the underlying map. Each cell of the grid can then be considered as a different stream. Our entire methodology is fully compatible with this setup.

## 3. COMBINATORIAL PATTERNS

In this section we introduce STComb, an approach for identifying combinational spatiotemporal patterns. These patterns are defined as combination of a temporal interval and a set of streams, where each stream originates from a different location.

This approach builds upon our previous work [14] on temporal bursts. Given a single stream of documents and a term $t$, we showed how we can extract, in linear time, the set of non-overlapping *bursty temporal intervals*. Given the sequence $Y_t = y_1, y_2, ... y_{|Y_t|}$ of frequency measurements for a term $t$, we defined the temporal burstiness of a given interval $I = Y_t[l:r]$ as a follows:

$$\mathcal{B}_T(I) = \left( \frac{\sum_{i=l}^{r} Y_t[i]}{\sum_{j=1}^{|Y_t|} Y_t[j]} - \frac{|I|}{|Y_t|} \right) \quad (1)$$

Note that the temporal burstiness $\mathcal{B}_T(I)$ of an interval $I$ is always in $[0, 1]$. This definition is based on the well-known discrepancy paradigm [5]. Based on this formalization, we presented an algorithm for the identification of high-burstiness intervals. Here, we extend this work in order to efficiently deal with *multiple streams* from different geographical locations. First, we use our previous method [14] to independently extract the sets of bursty temporal intervals for each stream. While we use our own formalization of temporal burstiness in our experiments, our methodology is compatible with any framework that reports non-overlapping bursty intervals.

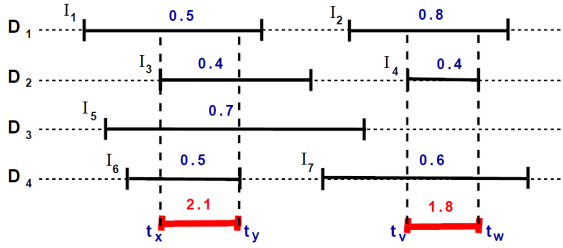

**Figure 2: Examples of bursty temporal intervals for 4 document streams $D_1, D_2, D_3$ and $D_4$**

Note that, since the intervals reported for each stream are strictly non-overlapping [14], overlap can only exist between intervals from different streams. Each segment that exists in the overlap of multiple intervals represents a spatiotemporal pattern, defined by the timeframe spanned by the segment and the set of the corresponding locations. Figure 2 shows examples of bursty temporal intervals for four document streams $D_1, D_2, D_3$ and $D_4$. In this example, two intervals $I_1$ and $I_2$ have been identified for $D_1$, with their respective (temporal) burstiness scores being 0.8 and 0.5.

Let $\mathcal{I}$ be the complete set of temporal intervals reported from all the document streams. Then, the problem of identifying spatiotemporal patterns is translated into finding subsets of overlapping intervals. A subset $\mathcal{I}' \subseteq \mathcal{I}$ is eligible only if all the intervals it includes share a common segment. Formally, $\mathcal{I}'$ is eligible if:

$$\bigcap_{I \in \mathcal{I}'} I \neq \emptyset \quad (2)$$

In Figure 2, we have $\mathcal{I} = \{I_1, I_2, I_3, I_4, I_5, I_6, I_7\}$. In this case, the subsets $\{I_1, I_3, I_5, I_6,\}$ and $\{I_2, I_4, I_7\}$ are eligible. On the other hand, the subset $\{I_1, I_4, I_6\}$ is not eligible.

To aid us in our analysis, we define $\mathcal{U}$ to be the universe of all eligible subsets of $\mathcal{I}$. First, we formally define the problem of finding the single highest-scoring subset of intervals:

PROBLEM 1. **Highest-Scoring Subset (HSS):** *Let $\mathcal{U}$ be the set of eligible subsets, and let $\mathcal{B}_T(I)$ return the temporal burstiness score of a given interval $I$. Then, we want to find the subset $\mathcal{I}^* \in \mathcal{U}$ such that:*

$$\mathcal{I}^* = \underset{\mathcal{I}' \in \mathcal{U}}{\operatorname{argmax}} \sum_{I \in \mathcal{I}'} \mathcal{B}_T(I) \quad (3)$$

Solving the HSS problem gives us the single highest scoring spatiotemporal pattern. Toward the end of this section we discuss how we can retrieve multiple high-scoring patterns. Note that any subset of intervals $\mathcal{I}' \in \mathcal{U}$ can be trivially converted into a (combinatorial) spatiotemporal pattern. First, by the definition of $\mathcal{U}$, each interval in $\mathcal{I}'$ comes from a different stream. Hence, all the streams that are represented in $\mathcal{I}'$ compose the set of streams of the pattern. Further, the timeframe of the pattern is defined as the common segment of all the intervals in $\mathcal{I}'$. Finally, the burstiness score is equal to $\sum_{I \in \mathcal{I}'} \mathcal{B}_T(I)$.

In the example of Figure 2, the highest scoring subset is $\{I_1, I_3, I_5, I_6\}$, which gives us the top spatiotemporal pattern. The set of streams included in the pattern is $\{D_1, D_2, D_3, D_4\}$. Further, the timeframe of the pattern is defined by the common segment of the intervals, spanning from timestamp $t_x$ to timestamp $t_y$ in the figure. Finally, the burstiness of the pattern is 2.1, equal to the cumulative burstiness of the included intervals.

Before we present our solution to the HSS problem, we state the following lemma, to aid us in our analysis:

LEMMA 1. *Given a set $\mathcal{I} = \{I_1, ..., I_m\}$ of 1-D intervals on the real line, the following two statements are equivalent:*

$$\bigcap_{I \in \mathcal{I}} I \neq \emptyset \quad (4)$$

$$I_i \cap I_j \neq \emptyset, \; \forall (I_i, I_j) \in \mathcal{I}, 1 \leq i, j \leq m \quad (5)$$

Lemma 1 simply states that if $m$ intervals have a non-empty intersection, then each pair of intervals must also have a non-empty intersection. The proof is trivial and is omitted for lack of space. Given Lemma 1, we can now state the following Proposition:

PROPOSITION 1. *The HSS problem is equivalent to the Maximum-Weight Clique Problem for Interval Graphs (MWCI)*

A detailed proof of proposition 1 can be found in Section A.1 of the Appendix. An instance of the Maximum-Weight Clique (MWC) problem consists of an undirected graph $G(V, E)$ and a vertex weight $w(v), \forall v \in V$. Given a constant $K$, the decision version of the MWC problem asks whether there exists a clique $V^* \subseteq V$, so that $\sum_{v \in V^*} w(v) \geq K$.

Proposition 1 refers to a specialized formulation of this problem, focusing exclusively on *Interval* Graphs (MWCI). An interval graph is the intersection graph of a set of intervals on the real line. It has a vertex for each interval in the set, and an edge between every pair of vertices corresponding to two intersecting intervals.

Proposition 1 allows us to use any known algorithms for the MWCI to solve HSS. In addition, while MWC is known to be NP-Complete [1], MWCI is solvable in polynomial time [8]. In our experiments, we use the algorithm described in [8], which returns the single highest-scoring clique in $O(n \log n)$ time. We refer to this algorithm as maxClique.

**Getting Multiple Patterns:** In order to obtain multiple non-overlapping patterns we can iteratively apply maxClique, removing each time the intervals included in the maximum clique. Allowing overlap would inevitably lead to uninformative results, obtained by trivially modifying other high-scoring cliques. Nonetheless, one can alternatively use any of the available algorithms for the enumeration of overlapping maximal cliques for interval graphs [32].

## 4. REGIONAL PATTERNS

The STComb algorithm presented in the previous section disregards the spatial proximity of the streams, and is thus inapplicable if one wishes to capture the spatial locality of bursts. In addition, STComb is not customized for streaming data, since it needs to recompute the set of cliques every time new information arrives. Next, we describe an online approach, called STLocal, that addresses these issues. By considering the geographical proximity among streams, we can evaluate the spatial extent of a term's burstiness pattern. Conceptually, we are looking for *bursty regions of the map*, instead of *arbitrary sets of bursty streams*.

First, we examine the simple case where we are given $D_x[i]$: the set of documents received from a single stream $D_x \in \mathcal{D}$ at timestamp $i$. We then extend our approach to

838

deal with a snapshot *of the entire collection* (i.e. all available streams), taken at some fixed point in time. Finally, we address the streaming scenario, where a new snapshot is added at every new timestamp.

***Single Data Stream:*** We model the spatiotemporal burstiness of terms by using the formal concept of *Discrepancy*. Discrepancy Theory has different formalizations and applications in several fields [5] and is generally used to describe the deviation of an *observed* situation from the *expected* baseline. Next, we use this paradigm to model the burstiness of a given term $t$: let $D_x[i]$ represent the set of documents that arrived from a stream $D_x \in \mathcal{D}$ at timestamp $i$. Then, given a term $t$, let $D_x[i][t]$ return the total frequency of $t$ in the documents included in $D_x[i]$. Formally:

$$D_x[i][t] = \sum_{d \in D_x[i]} freq(t,d) \qquad (6)$$

$D_x[\cdot][\cdot]$ can be visualized as a 2-D matrix, where rows correspond to timestamps and columns to terms. Then, $D_x[i][t]$ represents the *observed* frequency for $t$ on timestamp $i$.

Following the typical Discrepancy paradigm, we now define $E_x[i][t]$ to be the *expected* frequency of $t$ with respect to stream $D_x$ at timestamp $i$. This allows us to identify and evaluate frequency bursts by measuring the extent to which the *observed* frequency surpasses the *expected* baseline. The nature of an appropriate baseline depends on the domain of the application and the specifics of the data: $E_x[i][t]$ can be taken to be equal to the average observed frequency of $t$ in $D_x$, taken over all the snapshots collected before timestamp $i$. Alternatively, one can focus only on the most recent measurements. Finally, data from previous timeframes can also serve as a baseline, if available. For example, the expected frequency of a given term $t$ in the news from *San Francisco* on Dec-25-09 can be computed as the average daily frequency of the term, as computed over the measurements taken during the Dec. of previous years. Further discussion on possible baselines can also be found in our previous work on the temporal burstiness of terms [14].

Formally, we define the burstiness of a given term $t$ with respect to a data stream $D_x \in \mathcal{D}$ at timestamp $i$ as follows:

$$\mathcal{B}(t, D_x[i]) = D_x[i][t] - E_x[i][t] \qquad (7)$$

***Snapshot of the Entire Collection:*** Next, we discuss how we process an entire snapshot of the considered collection. A snapshot $\mathcal{D}[i] = \{D_1[i], D_2[i], ..., D_n[i]\}$ of a spatiotemporal collection $\mathcal{D}$ consists of the document-sets reported by *all* the streams at a single timestamp $i$. Our STLocal algorithm considers the spatial proximity of the streams in the 2-dimensional space: we want to find *regions* that are bursty with respect to a given term $t$. The burstiness of a region is based on the streams that originate within its area. Ideally, we could afford the flexibility of looking for regions of arbitrary shapes. However, this would dramatically increase the computational cost. Therefore, we focus on regions that can be represented by axis-oriented rectangles, allowing, as we show later, for a polynomial-time solution of the problem. By allowing rectangles of arbitrary size, we can capture interesting patterns on the 2-D map, while achieving an acceptable computational cost. A rectangle may contain multiple streams, depending on its size and location on the map. In the example of Figure 1, the rectangular area in north Africa includes streams $D_5$ and $D_6$.

We define the rectangle score (*r-score*) of a rectangle $R$ with respect to a term $t$ at a given timestamp $i$ as the sum of the respective burstiness values of the streams that fall within $R$. Formally:

$$\textit{r-score}(R, i, t) = \sum_{D_x \in R} \mathcal{B}(t, D_x[i]) \qquad (8)$$

where $\mathcal{B}(t, D_x[i])$ is as defined in Eq. 7. We can now formalize the notion of *Bursty Rectangles* as follows:

DEFINITION 1. [***Bursty Rectangles***]: Given a term $t$ and a snapshot $\mathcal{D}[i]$ of a spatiotemporal collection $\mathcal{D}$, we define as *Bursty Rectangles* the set of **non-overlapping** rectangles, for which $\textit{r-score}(\cdot, i, t) > 0$.

Positive-scoring rectangles represent regions where the overall observed frequency was higher than the expected one. The no-overlap constraint bounds the number of rectangles to at most $n = |\mathcal{D}|$. It also eliminates trivial results, produced by slightly modifying other high-scoring rectangles. In some cases, a higher *r-score* can be achieved by expanding the rectangle to include more streams, even if it means also including some non-bursty streams. Our approach automatically determines whether a set of streams should be included in a single rectangle, or if reporting a set of (two or more) smaller rectangles would benefit the *r-score*.

In Algorithm 1, we give the pseudocode of R-Bursty, an optimal algorithm that finds *all* non-overlapping *bursty rectangles* with a positive *r-score*. The R-Bursty algorithm uses as a module the polynomial algorithm proposed by Dobkin et al. [5] to find the single axis-oriented rectangle with the maximum bichromatic discrepancy in a 2-D setup. We refer the reader to the original paper for more details on this algorithm. A formal analysis of the complexity of R-Bursty can be found in Section A of the Appendix.

***Streaming Data:*** The R-Bursty algorithm provides us with the set of bursty rectangles for a single snapshot of the collection. As new snapshots arrive in a streaming fashion, we want to aggregate the consecutive rectangle-sets, in order to identify extended periods of time when particular regions of the map displayed bursty behavior. To assist us with the analysis, we define the concept of the *spatiotemporal window* $w = (R, [a:b])$, consisting of an axis-oriented rectangle $R$ and the timeframe $[a:b]$. Geometrically, a spatiotemporal window $w$ can be represented as a hyper-rectangle in 3-D space. Figure 3 shows 3 different examples of spatiotemporal windows, $w_1, w_2$ and $w_3$, on a $60 \times 40$ map. Window $w_1$ corresponds to the rectangle $R$ on the map, and spans the timeframe between 3 and 8. Also, observe that $w_2$ and $w_3$ correspond to the same rectangle, even though they span different timeframes. Given a term $t$, we define the burstiness (*w-score*) of a spatiotemporal window $w = (R, [a:b])$ with respect to a term $t$ as follows:

$$\textit{w-score}(w, t) = \sum_{i=a}^{b} \textit{r-score}(R, i, t) = \sum_{i=a}^{b} \sum_{D_x \in R} \mathcal{B}(t, D_x[i]) \qquad (9)$$

## Algorithm 1 R-Bursty

**Input:** term $t$, snapshot $\mathcal{D}_i$ of a spatiotemporal collection $\mathcal{D}$
**Output:** All **non-overlapping** rectangles in $\mathcal{D}_i$ that have $r\text{-}score(\cdot, i, t) > 0$.

1: Run the algorithm in [5] to retrieve $R_{max}$, the rectangle in $\mathcal{D}_i$ with the highest $r\text{-}score$.
2: Report $R_{max}$ and set $\mathcal{B}(t, \mathcal{D}_x) = -\infty, \forall \mathcal{D}_x \in R_{max}$
   ( We set the scores of the streams within $R_{max}$ to $-\infty$ to eliminate overlap among the reported rectangles).
3: Repeat the process from the step (1), until the $r\text{-}score$ of the retrieved rectangle is less or equal to zero.

## Algorithm 2 STLocal

**Input:** Spatiotemporal collection $\mathcal{D}$
**Output:** Set of Maximal Windows $\mathcal{W}_t$ for every term $t$
1: $i \leftarrow 0$ // Timestamp Counter
2: Initialize $\mathcal{S}_t \leftarrow \emptyset, \mathcal{W}_t \leftarrow \emptyset$ for every term $t$
   ($\mathcal{S}_t$ contains a sequence of snapshots for every rectangle)
3: **while** Stream is open **do**
4:   $i \leftarrow i + 1$
5:   **for** each term $t$ **do**
6:     $\mathcal{R} \leftarrow$ R-Bursty$(\mathcal{D}_i, t)$
7:     $\mathcal{S}_t \leftarrow \mathcal{S}_t \cup \{\text{new sequence } S : \forall R \in \mathcal{R}\}$
8:     **for** ( each sequence $S \in \mathcal{S}_t$ ) **do**
9:       $S.\text{add}(r\text{-}score(R_S, i, t))$
10:      $\mathcal{W}_t \leftarrow \mathcal{W}_t \cup \text{GetMax}(S)$
11:      **if** ($S.total < 0$) **then**
12:        Remove $S$ from $\mathcal{S}_t$

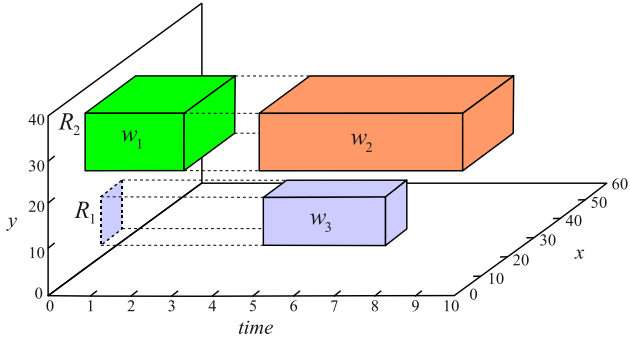

**Figure 3:** Examples of Spatiotemporal Windows. Both windows $w_2$ and $w_3$ correspond to the same rectangular region ($R_2$), while $w_3$ corresponds to a different region ($R_1$).

While Eq. 9 allows us to evaluate the score of any given window, it is computationally intractable to consider all possible windows in order to find those with the highest scores. Before we discuss how we tackle this problem, we formalize the concept of a *maximal* spatiotemporal window:

DEFINITION 2. [***Maximal Spatiotemporal Window***]: Given two windows $w = (R, [a:b])$ and $w' = (R', [a':b'])$, we say that $w'$ is a *sub-window* of $w$ if $w'$ is completely contained in $w$ (in terms of both space and time, i.e., $R' \subseteq R$, $b' \leq b$ and $a' \geq a$). Thus, $w$ is then considered a *super-window* of $w'$. Then, a window $w$ is considered *maximal* if and only if there exist no super-windows of $w$ that have a higher $w\text{-}score$ than it does.

The above definition allows us to find high-scoring, non-overlapping windows, while avoiding trivial occurrences, such as windows that trivially differ from other high-scoring windows or over-extended windows that may in fact contain higher-scoring candidates. Therefore, in our context, the maximal window represents a meaningful and informative spatiotemporal pattern. Given this concept, we formalize the Bursty Source Patterns problem by mapping it to the problem of finding the set of *Maximal Windows*. Given a spatiotemporal collection $\mathcal{D}$ and a term $t$, we want to find the set of maximal spatiotemporal windows $\mathcal{W}_t$.

A positive score means that, within the region covered by the window, the observed frequency of the term was higher than the expected one. In the context of streaming data from multiple streams, computing and maintaining $W_t$ is a non-trivial task. In Algorithm 2, we present STLocal, an efficient algorithm that finds the *Maximal Windows*.

**A walkthrough of the STLocal Algorithm:** Given a spatiotemporal collection $\mathcal{D}$, the algorithm maintains a set of maximal windows $\mathcal{W}_t$ for each term $t$. For every new snapshot $\mathcal{D}_i$, we use R-Bursty to identify the respective set $\mathcal{R}$ of bursty rectangles (Line 6). For each term $t$, the algorithm maintains a set of sequences $\mathcal{S}_t$, corresponding to the set of rectangles that have been identified as bursty. For every sequence $S \in \mathcal{S}_t$, let $R_S$ represent the region of the map that $S$ corresponds to. Every time a new snapshots arrives, the $r\text{-}score$ of $R_S$ is computed and appended to $S$ (Line 9).

Given such a sequence of real values, we need an online process able to compute and maintain $\mathcal{W}_t$. For this, we employ the online algorithm presented by Ruzzo and Tompa [21], which we refer to as GetMax. Given a sequence of real values, GetMax identifies all the maximal segments (i.e. contiguous subsequences) in linear time. A description of GetMax can be found in Section C of the Appendix. Each maximal segment corresponds to a maximal window. In Line 10, GetMax is used to update the set of maximal windows $\mathcal{W}_t$ for the term. In practice, the algorithm is not re-applied to the entire sequence every time a new score is appended. Instead, it maintains the processed sequence and updates the set of maximal windows every time a new value arrives.

In Lines 11-12, we eliminate regions (and their respective sequences) that should no longer be considered bursty. When $S.total$ (the sum of all the scores in $S$) becomes negative, $S$ cannot be a part of a maximal segment and it is safe to remove it from $\mathcal{S}_t$. Conceptually, no maximal segment can have a suffix of $S$ as its prefix. The proof is straightforward and is omitted for lack of space.

The algorithm is polynomial in the number of streams. A formal complexity analysis of STLocal can be found in Section A of the Appendix.

**Discussion on proximity:** STLocal captures patterns that correspond to 2D *rectangles* of arbitrary size. This choice was principally motivated by the need to combine flexibility with low (polynomial) computational cost. Further, it is possible for a bursty rectangular region to contain a small number of non-bursty streams. This number is bound to be small, otherwise one could split the region and obtain higher-scoring rectangles that do not include such streams. Nonetheless, it is computationally trivial to remember, and ultimately exclude, such "false positives" for each pattern.

## 5. FINDING BURSTY DOCUMENTS

Next, we show how we can use spatiotemporal patterns to retrieve documents that are relevant to a user's query and also discuss events with a high spatiotemporal impact. We refer to these documents as *bursty documents*. Even though our search engine is compatible with both regional and combinatorial patterns, it only handles one type at a time (i.e. a separate instance is required for each type).

On a high-level, our search engine considers two factors in the evaluation of a given document: **(1)** the relevance of the document to the user's query; and **(2)** the document's burstiness, as captured in its overlap with the mined spatiotemporal burstiness patterns. Formally, given a query of terms $q$, the score of a document $d$ is computed as follows:

$$score(q,d) = \sum_{t \in q} relevance(d,t) \times burstiness(d,t) \quad (10)$$

Here, *relevance(d,t)* is the relevance of document $d$ with respect to term $t$. This can be implemented as any normalized version of *freq(t,d)*, i.e. the number of occurrences of $t$ in $d$. The best choice depends on the particular nature of the considered documents. In our own experiments, we found that using $\log(freq(t,d+1))$ yielded the best results. Further, *burstiness(d,t)* is the burstiness of document $d$ with respect to term $t$. Let $\mathcal{P}_t$ be the set of patterns extracted for a given term $t$. Recall that both types of spatiotemporal patterns discussed in this paper (combinational and regional) include a timeframe and a set of streams. In addition, each document $d$ arrives from a single stream at a specific point in time. We say that $d$ *overlaps* with a pattern $P$ if both its stream of origin and its timestamp are included in $P$. Formally, let $\mathcal{P}_{t,d} \subseteq \mathcal{P}_t$ be the subset of patterns for term $t$ that overlap with a given document $d$. Then, we define the burstiness of $d$ with respect to $t$ as follows:

$$burstiness(d,t) = \begin{cases} f(\mathcal{P}_{t,d}) & \text{if } \mathcal{P}_{t,d} \neq \emptyset \\ -\infty & \text{otherwise} \end{cases} \quad (11)$$

where $f(\mathcal{P}_{t,d})$ can be any function of the scores of the patterns in $\mathcal{P}_{t,d}$. For example, $f(\cdot)$ can return the maximum, minimum or median such score. An aggregate function that considers all the scores, such as the average, can also be applied. In our own experiments, we found that using the maximum score over all the patterns included in $\mathcal{P}_{t,d}$ yielded the best results. Given Eq. 10, we can now formulate the *Bursty Documents* problem:

PROBLEM 2. [**Bursty Documents**]: Given a set of streams $\mathcal{D}$ and a query of terms $q = \{t_0, t_1, ...\}$, we want to find the $k$ documents from $\mathcal{D}$ with the highest burstiness, i.e. those $k$ documents that maximize Eq. 10.

The problem can now be addressed via standard information retrieval techniques. An inverted index is first built, mapping each term to the documents that include it, ranked by their respective scores. The popular Threshold Algorithm (TA) [6] for top-k evaluation can then be applied to retrieve the top documents for any given multi-term query.

## 6. EXPERIMENTAL EVALUATION

In this section we present the experimental evaluation of our methodology for the identification and utilization of spatiotemporal patterns. Our evaluation tests our methodology in terms of both the quality of the reported results and the computational efficiency. We begin in Section 6.1, where we describe the datasets used. In Section 6.2, we demonstrate the effectiveness of our two frameworks for the identification of bursty spatiotemporal patterns. We also discuss their differences, as they emerge from the experimental findings. The mined patterns are evaluated in terms of both space (i.e. the streams that they include) and time (i.e. the timeframe that they span). In Section 6.3 we evaluate and compare the two frameworks in the context of the Bursty Documents Problem. We complete our experiments in Section 6.4, where we evaluate and discuss the performance of the two algorithms.

### 6.1 Datasets

**Topix Dataset:** For lack of an openly available dataset of proper sequences (with consecutive timestamps) of documents from different geographic locations, we composed a corpus of articles from Topix.com, which hosts news-stories from different countries around the world. This dataset contains 305,641 articles, where the vast majority of them come from local news sources from 181 different countries, posted between Sep-08 and Jul-09. To project the sources' locations on the 2-D plane, we use Multidimensional Scaling [30] given the pair-wise geographical distances of sources.

**Major Events List:** We composed a list of influential real-life events that took place during the timeframe spanned by the dataset. The events were taken from WikiPedia.com which maintains a list of major events for every calendar year. We identify three loosely-defined categories of events in the list: events with a significant global impact (events 1–6), major events that were reported in a significant number of countries (7–12) and events with a more localized impact (13–18). A short description of the selected events is given in Table 9 of the Appendix. Each event was shown to a human annotator, who was instructed to provide the query that she would submit to a search engine, in order to retrieve information on that event (second column of Table 9).

**Artificial Data:** Our work is the first to study spatiotemporal term-burstiness. Thus, the lack of real datasets with a provided ground truth can be a problem for the evaluation process. To address this, we apply two different generators for appropriate spatiotemporal data, which we refer to as DISTGEN and RANDGEN. DISTGEN emulates a realistic scenario for the creation of spatiotemporal patterns, while RANDGEN takes a purely randomized approach. The generators are described in detail in section B of the Appendix.

### 6.2 Spatiotemporal Pattern Evaluation

In this experiment we use our two approaches to retrieve the top-scoring burstiness pattern, given each of the queries in Major Events List. Table 1 shows the number of countries included in the top combinatorial pattern by STComb and the top regional pattern by STLocal. For STComb, we also report the number of countries included in the Minimum Bounding Rectangle (MBR) of the set of countries included in the top clique. This essentially gives us the number of streams that were included in the region delimited by the included streams.

Table 1 provides valuable insight on the behavior of the two algorithms. For events with a significant global impact (e.g. the death of singer Michael Jackson or the global financial crisis), both STLocal and STComb report large spatiotemporal patterns, covering the majority of the available data sources. For the events of the middle tier (e.g. the acts

841

Table 1: Top-Scoring Bursty Source Patterns.

| # | Query | # countries in STLocal | # countries in STComb | # countries in MBR |
|---|---|---|---|---|
| 1 | Obama | 176 | 136 | 181 |
| 2 | financial crisis | 113 | 159 | 181 |
| 3 | Jackson | 132 | 151 | 181 |
| 4 | terrorists | 98 | 126 | 167 |
| 5 | swine | 174 | 157 | 181 |
| 6 | earthquake | 17 | 81 | 171 |
| 7 | gaza | 174 | 116 | 179 |
| 8 | ceasefire | 36 | 52 | 156 |
| 9 | Yemenia | 19 | 21 | 125 |
| 10 | piracy | 24 | 39 | 174 |
| 11 | Air France | 50 | 67 | 179 |
| 12 | bush fires | 3 | 30 | 168 |
| 13 | Nkunda | 30 | 2 | 118 |
| 14 | Vieira | 15 | 22 | 114 |
| 15 | Tsvangirai | 4 | 24 | 123 |
| 16 | Rajoelina | 4 | 30 | 154 |
| 17 | Fujimori | 5 | 19 | 158 |
| 18 | Zelaya | 26 | 55 | 171 |

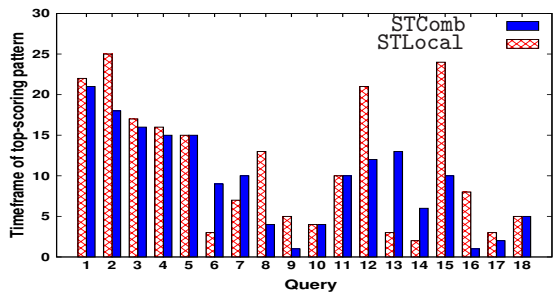

Figure 4: Timeframe length of the top pattern for queries in the Major Events List.

of piracy in Somalia), the results of the two approaches begin to differ, with STComb generally including more countries in the top pattern. Finally, This difference becomes even more apparent for events with a more localized impact (e.g. the inauguration of M. Tsvangirai as the new Prime Minister of Zimbabwe). For such events, STLocal reports small patterns that focus around the event's source. On the other hand, STComb reports larger patterns with countries from around the globe. This behavior can be explained by the fact that STLocal is bounded by the geographical proximity among the streams. On the other hand, STComb focuses exclusively on the maximization of burstiness, resulting in larger patterns with sources from arbitrary locations on the map. This is also demonstrated by the very large sets of countries included in the MBR of the reported patterns.

### 6.2.1 Timeframe Evaluation

We complete our analysis with the evaluation of the timeframes of the reported patterns, plotted in Figure 4: each pair of bars corresponds to a query, following the same order as in Table 1. The left bar of each pair represents the timeframe spanned by the regional pattern given by STLocal. The right bar represents the timeframe of the combinatorial pattern given by STComb. The y-axis represents the length of the timeframe in weeks. For most queries, the two approaches report timeframes of a similar length. For some cases, however, we observe that STLocal reports longer timeframes. This behavior is observed for events that remain in the local spotlight even after the event has faced in locations further from the source.

In conclusion, our experimental findings verify that the two proposed algorithms fulfill the purposes for which they were designed: STLocal can track the spatiotemporal impact of events, which is especially meaningful for events that affect specific regions. On the other hand, STComb can be used to identify all the locations affected by an event, regardless of their geographical coordinates.

### 6.2.2 Pattern Retrieval on Artificial data

Next, we use artificial data to evaluate our approaches in the task of retrieving spatiotemporal patterns. For this experiment, we generated two datasets of streaming data. Each dataset was injected with 1000 spatiotemporal patterns. The length of the timeline was set to 365 (to emulate the number of days in a calendar year), and the number of considered terms was set to 10000. The DISTGEN process was used to generate the patterns for the first dataset, and the RANDGEN process was applied for the second one (see Section B of the Appendix for details on the generators). We then used STLocal and STComb to retrieve the timeframe and the included streams of the injected patterns. For each pattern, we first check how each approach performs in retrieving the included streams. Given the set of retrieved streams $Y$ and the actual set of streams included in the pattern $Y'$, we report the Jaccard Coefficient of the two values, defined as $\frac{|Y \cap Y'|}{|Y \cup Y'|}$. We refer to this quantity as *JaccardSim*. Higher values are desirable. In addition, we check how each approach fared in finding the first and last timestamp of each pattern's timeframe. Let $i$ be the actual value of the first timestamp of the pattern, and let $i'$ be the respective value reported by the approach. Then, we report the *Start-Error* $|i - i'|$. We similarly compute the *End-Error* for the pattern's last timestamp. Finally, we report the average for each measure, computed over all the patterns in each dataset.

**A baseline:** For this task, we compare our methods with the following baseline. First, we compute the burstiness of a given term $t$ with respect to a stream $D_x$ at timestamp $i$ by using Eq. 7. This gives us a sequence of scores for each stream. Positive (bursty) scores are replaced with '1', while negatives ones with '0'. Contiguous segments of ones represent timeframes of bursty activity for the stream. To allow for some gap in the segments we replace any contiguous segment of zeros that has length less than $\ell$ (and is not in the beginning or end of the sequence) with an equal segment of ones. This gives us the final set of bursty intervals for each source. Given a random order of the streams, let $\mathcal{I}$ be the set of intervals of the first stream. Then, for each interval $I$ in the interval-set of the next stream, check if there is an interval $I' \in \mathcal{I}$ so that $\frac{|I' \cap I|}{|I' \cup I|} \geq \delta$. If such an interval exists, $I$ and $I'$ are merged, and $I' \cap I$ replaces $I$ in $\mathcal{I}$. The process continues until all streams have been processed. We refer to this method as Base. For our experiments, we tune both the $\ell$ and $\delta$ parameters to yield the best results.

The results of the experiment are shown in in Table 2. Each row holds the values achieved by an approach for JaccardSim, Start-Error and End-Error, for one of the two datasets (i.e. the one generated via DISTGEN or RANDGEN). As can be seen from the table, STLocal produces great results for the realistic DISTGEN scenario, which emulates the spatiotemporal patterns of real events. The high value of of JaccardSim demonstrates that this approach can accurately identify the streams that are affected by the pattern.

The timeframe is also accurately gauged, as demonstrated by the low Start-Error and End-Error values. For the RANDGEN scenario, STLocal was again accurate in identifying the timeframe of the patterns. The method also achieves a satisfactory JaccardSim value of 0.72, although still lower than in the case of DISTGEN. This can be explained by the lack of spatial locality in the patterns generated by RANDGEN.

Table 2: Spatiotemporal pattern retrieval.

|  | JaccardSim | Start-Error | End-Error |
| --- | --- | --- | --- |
| STLocal | | | |
| DISTGEN | 0.88 | 6.4 | 9.8 |
| RANDGEN | 0.72 | 16.6 | 15.2 |
| STComb | | | |
| DISTGEN | 0.69 | 6.2 | 10.6 |
| RANDGEN | 0.91 | 12.6 | 9.3 |
| Base | | | |
| DISTGEN | 0.34 | 44.4 | 45.1 |
| RANDGEN | 0.52 | 30.3 | 29.8 |

Contrary to STLocal, STComb achieves higher values for the RANDGEN scenario. Since this approach is not restricted by spatial locality, it is better suited for identifying patterns of arbitrary nature. On the other hand, this makes STComb less effective in the DISTGEN scenario, where the spatial locality of the more realistic patterns is influential. A detailed study of the results, revealed that, for DISTGEN, STComb would sometimes overlook bursty streams around the source of the considered patterns, thrown off by irrelevant bursty behavior from arbitrary locations on the map.

Finally, Base was consistently outperformed by the other two approaches, achieving worse values for all considered measures. In addition to verifying that the retrieval task is non-trivial, this hints at the advantage of simultaneously handling both the spatial and temporal dimensions of textual data, as performed by STLocal and STComb.

### 6.3 Bursty Documents Evaluation

In this experiment, we evaluate the two proposed approaches in the context of the Bursty Documents problem. Given the set of events from Major Events List, and their respective queries, we use STLocal and STComb to retrieve the top-10 documents for each event. The retrieval process is performed exactly as described in Section 5. The retrieved documents are then given to a human annotator, who marks each of them as "relevant" or "not relevant" to the event. This allows us to evaluate the *precision* of the two approaches.

We compare the results with the search engine described in [14], which focuses exclusively on *temporal* term burstiness. We refer to this approach as TB. Since this approach disregards the origin of each document, the streams from the various countries were merged to a single stream.

The three approaches consistently reported high precision, as shown in Table 3. STLocal was perfect for all queries and STComb for all except one ($Q_{13}$, with 80% precision). TB had a few false positives for the events in the 3rd category (i.e. the ones with a more localized impact), with an average of 80% precision. This can be explained by the fact that TB focuses only on the global maximization of the temporal burstiness, assuming a single source. Therefore, TB can be less sensitive to events with a more limited, localized impact. A characteristic example is the query "earthquake": all 10 documents returned by STLocal discussed the 2009 Costa Rica *Cinchona Earthquake*. This was anticipated, since the algorithm considers the geographical locations and proximity of the sources on the map.

Table 3: Precision in top-10 documents.

| # | Query | TB | STLocal | STComb |
| --- | --- | --- | --- | --- |
| 1 | Obama | 1.0 | 1.0 | 1.0 |
| 2 | financial crisis | 1.0 | 1.0 | 1.0 |
| 3 | terrorist | 1.0 | 1.0 | 1.0 |
| 4 | Jackson | 0.9 | 1.0 | 1.0 |
| 5 | swine | 1.0 | 1.0 | 1.0 |
| 6 | earthquake | 1.0 | 1.0 | 1.0 |
| 7 | gaza | 1.0 | 1.0 | 1.0 |
| 8 | ceasefire | 1.0 | 1.0 | 1.0 |
| 9 | Yemenia | 1.0 | 1.0 | 1.0 |
| 10 | piracy | 1.0 | 1.0 | 1.0 |
| 11 | Air France | 1.0 | 1.0 | 1.0 |
| 12 | bush fires | 1.0 | 1.0 | 1.0 |
| 13 | Nkunda | 0.7 | 1.0 | 0.8 |
| 14 | Vieira | 0.8 | 1.0 | 1.0 |
| 15 | Tsvangirai | 0.9 | 1.0 | 1.0 |
| 16 | Rajoelina | 0.7 | 1.0 | 1.0 |
| 17 | Fujimori | 0.8 | 1.0 | 1.0 |
| 18 | Zelaya | 1.0 | 1.0 | 1.0 |

Among the documents given by STComb, 3 were on the Sichuan earthquake in China, 3 were on an earthquake in Guerrero, Mexico and all the others discussed earthquakes from different countries across the world. Finally, for TB, 3 articles were on the same earthquake from Bulgaria, while all others discussed different locations.

To perform a more thorough analysis of the results, we calculate the similarity between their top-k sets (defined as the size of the overlap divided by 10). The observed similarity values where the following: STComb-TB: 0.61, STComb-STLocal: 0.58, and TB-STLocal: 0.67. The relatively low similarity values raise an interesting point: even though all three algorithms achieve an extremely high precision, their top-k sets can differ significantly. Therefore, by optimizing different facets of burstiness, the three approaches report diverse results and complement each other. Depending on the occasional application, one may choose to focus on a particular approach, or consider the rankings of all three approaches toward a ensemble method.

### 6.4 Performance Evaluation

As shown in Section A of the Appendix, the complexity of the STLocal algorithm is $O(|L|n^3 \log n)$, where $n$ is the number of data sources and $|L|$ the length of the stream (i.e. number of timestamps).

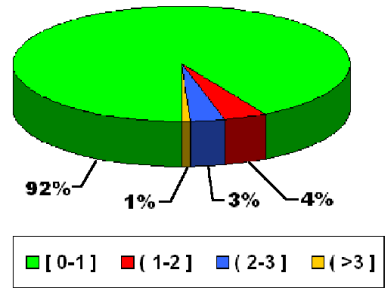

Figure 5: Distribution of the number of rectangles per timestamp for STLocal on the Topix dataset.

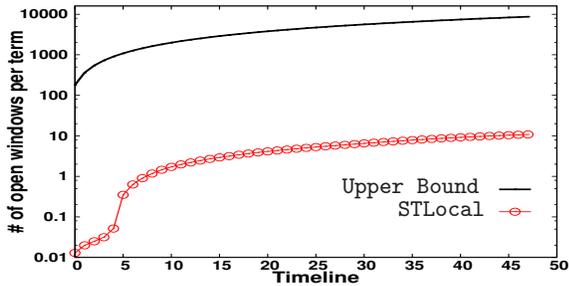

**Figure 6: Number of open spatiotemporal windows.**

This worst-case complexity assumes that, for a given term $t$, there exist $O(n)$ bursty rectangles in every 2-dimensional snapshot taken at a single timestamp. However, in practice, this number is a lot smaller than $n$. Next, we verify this on the Topix dataset, for which $n$ (number of countries) is equal to 181. First, we compute the average number of bursty rectangles reported for each term per timestamp. We then build a histogram of the computed population, visualized as a pie chart in Figure 5. The chart shows that, for the vast majority of terms (92%), the average number of rectangles per timestamp was between 0 and 1, far smaller than the 181 assumed by the worst-case scenario.

Another factor that affects the complexity of STLocal is the number of spatiotemporal windows that need to be maintained. The worst-case analysis assumes that, for a timeline of length $L$, this number is $O(n|L|)$ (i.e. $n$ new windows per timestamp). As we show using the Topix dataset, the number in practice is considerably smaller. For this dataset, $n|L|$ translates to a total of 181×48=8,688 distinct windows. The actual number of open windows per time instance, as reported by STLocal, is shown in Figure 6. The number shown is the average taken over all the terms in the collection. We also plot the worst-case number for each timestamp (181 for timestamp $i$=1, 362 for $i$=2, etc.). As can be seen by the figure, the number assumed by the worst-case scenario is several orders of magnitude larger than the one observed for real data, which peeks at around 10 open windows per term.

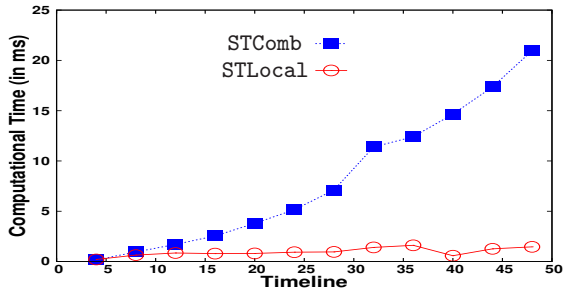

**Figure 7: Running time (ms) per timestamp.**

**Comparing STLocal and STComb in terms of speed:** Next, we compare the computational time required by our proposed algorithms to process the Topix dataset. Our experiment emulates the streaming scenario, i.e., we process the collection one timestamp at a time, in sorted order by timestamp. Since the processing of each term is independent for both algorithms, we report the average time required to process a single term in each timestamp. Figure 7 shows that STLocal clearly outperforms STComb. This was anticipated, since STLocal is an online algorithm, with the ability to update the information for each term, every time new data arrives. On the other hand, STComb needs to be re-applied to the entire updated dataset. STLocal consistently required times around 1ms, exhibiting great performance and scalability. That being said, it is important to note that the results for the STComb are encouraging: even when asked to process the entire stream, the algorithm required as little as 20ms per term. This illustrates the potential of STComb and motivates us to work on an online version of the algorithm.

### 6.4.1 Scalability on Artificial Data

We conclude our experiments by evaluating the scalability of STLocal and STComb on large artificial corpora, created via the DISTGEN process. We prefer DISTGEN over RANDGEN, since it emulates real spatiotemporal events and is thus more realistic. In particular, we use DISTGEN to generate datasets with a different number of streams $|\mathcal{D}|$, with $|\mathcal{D}| \in \{500, 1000, 2000, 4000, 8000, 16000, 32000, 64000, 128000\}$. The length of the timeline was set to 365, and each dataset was injected with 1000 patterns. The number of considered terms was set to 10000. We then use two approaches to retrieve the spatiotemporal patterns on each dataset. The results are shown in Figure 8.

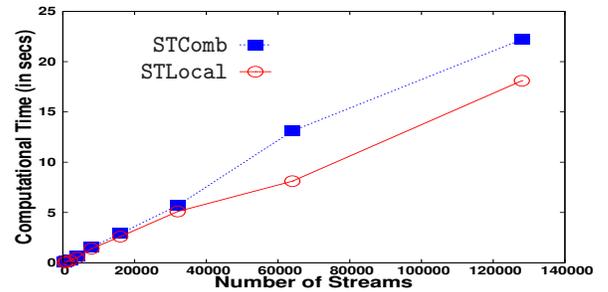

**Figure 8: Running time Vs Number of streams.**

The x-axis holds the number of streams, while the y-axis holds the respective computational time (in seconds) per term. As can be seen from the figure, both approaches scale almost linearly with the number of streams. Further, STLocal was consistently faster, even though the margin was not very wide. The scalability and online nature of STLocal make it an ideal candidate for streaming data.

## 7. RELATED WORK

To the best of our knowledge, ours is the first work to formalize the spatiotemporal burstiness of terms and utilize it toward an efficient search engine for the retrieval of documents on influential events. Nonetheless, we identify various works that have ties to our own.

A significant amount of work has been devoted to the evaluation and utilization of temporal burstiness [7, 11, 10, 31, 35]. Contrary to our own work, all these papers disregard the spatial dimension. Further, a number of works explore the spatiotemporal aspects of textual collections, albeit in a different context. Singh et al. [25] give a very brief overview of a pixel-based approach for the visualization of spatiotemporal events discussed in microblogging sites. Dalli [4] describes cpGeo, a system for large-scale analysis of blogs and online news. Yu and Mengham [20] explore the spatiotemporal dimension of the data to identify emerging trends. Tsoukatos and Gunopulos [27] focus on finding frequent patterns in spatiotemporal databases. Sankaranarayanan et al. [23] present a clustering technique that considers the users' locations and the content of the user's tweets on Twitter to identify topic locations.

Mei et al. [19] focus on spatiotemporal *theme mining* on blogs. Our problem setup is different in many ways. First, our formulation is dynamic in the way it considers both spatial and temporal information, since both the streams and timeframe of a pattern are identified automatically. In contrast, Mei et al. focus on finding the *life cycle* (timeframe) of a given theme, without reporting sets of bursty streams or regions on the map. Instead, the spatial dimension is only considered for a single fixed timestamp, for which they return the map of of distributions over all locations, for a given theme. In addition, they make no mention of document search, which is one of our primary contributions.

Sakaki et al. [22] use a binary SVM to classify Twitter feeds as relevant or non-relevant to a given event. They then perform a spatiotemporal analysis of the relevant feeds. This introduces the need for training data, contrary to our completely unsupervised approach. Bansal and Koudas [2] describe BlogScope, a search engine for blogs. In this work, the spatial dimension is considered in restricted manner, since users have to select a specific region of the map to view analytics. In an extension of this work [18], the goal is to find spatial bursts in a *fixed* temporal interval in a grid-based spatial layout. In contrast, our approach can simultaneously track spatial and temporal burstiness, and also has no requirement of a grid with fixed-cell size.

**Moving objects:** Even though our context differs in numerous ways, our work also has ties to the extensive literature on indexing, querying and mining moving objects through time and space. For instance, relevant research [9, 26] describes indexing techniques to efficiently find moving objects that match a user defined spatiotemporal predicate (e.g., "find all moving objects that are at most 1 mile to a given spatial location between June-11 and July-11"). Vieira et. al [29] propose a framework to query trajectories based on motion pattern queries. A pattern query is defined as a sequence of spatiotemporal predicates. Chen et. al [3] use a similar idea to retrieve similar trajectories, given a query set of spatial points. Another relevant line of work [33, 34, 17, 15], discusses mining techniques for discovering similar movement patterns and periodicities on historical spatiotemporal data. In the same domain, relevant papers [12, 28, 16] explore methods to discover group of trajectories that stay "close" together in space for a continuous timeframe. Our work differs significantly from those described above, since the queries we explore are purely textual and make no use of spatial or temporal predicates.

## 8. CONCLUSION

In this paper we formalized and evaluated the spatiotemporal burstiness of terms. We proposed two alternative approaches for mining spatiotemporal burstiness patterns, STComb and STLocal. The two approaches are complementary, providing valuable insight on spatiotemporal burstiness from different perspectives. We then showed how the mined patterns can be utilized toward an efficient document-search engine. Our engine returns documents on influential events with a major spatiotemporal impact. Finally, we demonstrated the efficacy and efficiency of our methods through a rigorous experimental evaluation on real data. In future work, we intend to extend our STLocal approach to handle geographical regions of arbitrary size, as opposed to the rectangular shapes that it can currently address. In addition, we intend to work on a purely online version of STComb, in order to enhance its applicability in real-life applications.

## 9. ACKNOWLEDGMENTS

This work was supported by the MODAP EU Project, the DISFER GGET Project, and by the NSF grants NSF-IIS:0910859 and NSF-IIS:1144158.

# APPENDIX

## A. ALGORITHM COMPLEXITIES

**Complexity of R-Bursty:** The complexity of the first step is $O(n^2 \log n)$ [5]. Since the number of non-overlapping rectangles is bounded by $n = |\mathcal{D}|$, the complexity of R-Bursty is $O(n^3 \log n)$. This polynomial cost becomes even more satisfactory if one considers that the number of streams $n$ is typically limited (i.e. in the tenths or hundreds).

**Complexity of STLocal:** Since each term is processed independently, the process can be easily parallelized. The complexity is then as follows: let $|L|$ be the length of the timeline spanned by our collection. STLocal applies the R-Bursty algorithm $|L|$ times, thus requiring $O(|L|n^3 \log n)$, where $n = |\mathcal{D}|$ is the number of streams. Further, the maximum number of sequences (i.e. bursty regions) that need to be maintained is $O(n|L|)$. As we show in the experiments, the actual number is a lot smaller, since bursty artifacts are, by definition, rare. By using GetMax, we can maintain each window in $O(|L|)$ time, for a total of $O(n|L|^2)$. Therefore, the overall complexity is $O(|L|n^3 \log n + n|L|^2) = O(|L|n^3 \log n)$.

## A.1 Proof of Proposition 1

We prove Proposition 1 by showing that the two problems are reducible to each other.

**From CB to MWCI:** Given a set $\mathcal{I}$ of temporal bursty intervals $\mathcal{I}$ on the real line, we create a node $v$ for every interval $I \in \mathcal{I}$ and add it to a set of nodes $V$. We also set $c\text{-}score(v) = \mathcal{B}_\mathcal{T}(I)$. We then add an edge between two nodes, if their corresponding intervals intersect. Let $E$ be the set of created edges. The above steps can be completed in $O(|V| + |E|)$ time. Now, Let $V^* \subseteq V$ be the Maximum-Weight Clique of graph $G(V, E)$ and let $\mathcal{I}^*$ be the set of intervals corresponding to the nodes in $V^*$. Since $V^*$ is a clique, we know that $I_i \cap I_j \neq \emptyset$, $\forall (I_i, I_j) \in \mathcal{I}^*$. Therefore, from Lemma 1, we know that $\bigcap_{I \in \mathcal{I}^*} I \neq \emptyset$, which is the 1st condition for the CB problem. Further, since $V^*$ is the **maximum-weight** clique, we know that the 2nd condition of the problem is also satisfied.

**From MWCI to CB:** We are given an Interval Graph $G(V, E)$, where each $v \in V$ is associated with a weight $c\text{-}score(v)$. We can obtain an instance of the CB problem by simply mapping $G(V, E)$ to its interval representation: we create a set of intervals $\mathcal{I} = \{I_v | \forall v \in V\}$, which we then place on the real line, so that two intervals intersect only if their corresponding nodes are connected by an edge in $E$. This can be done in linear time [24]. Each interval $I_v$ is assigned a burstiness score $\mathcal{B}_\mathcal{T}(I_v) = c\text{-}score(v)$.

Then, let $\mathcal{I}^* \subseteq \mathcal{I}$ be a solution of CB problem on $\mathcal{I}$ and let $V^* \subset V$ be the subset of nodes corresponding to the intervals in $\mathcal{I}^*$. From the 1st requirement of CB and Lemma 1, we know that $\exists (v_i, v_j) \in E$, $\forall v_i, v_j \in V^*$ and, therefore, $V^*$ is a clique. We also know that $V^*$ is a maximum-weight clique, since it maximizes $\sum_{v \in V^*} c\text{-}score(v) = \sum_{I \in \mathcal{I}^*} \mathcal{B}_\mathcal{T}(I)$.

## B. ARTIFICIAL DATA GENERATION

The input to the generators consists of the length of the considered timeline $T$, the number of streams $|\mathcal{D}|$, the number of considered terms, and the number of spatiotemporal patterns to be generated.

The first step is the creation of the frequency streams. Each stream is populated as follows. First, a random frequency is selected for each of the timestamps in the timeline. This value is randomly sampled based on an exponential distribution. This process generates the typical frequency of terms (i.e. not due to relevant events). Our experiments on the Topix dataset verified that the exponential distribution is a good fit for this task.

The next step is pattern generation. A pattern is generated as follows: first, the term that is to exhibit the pattern is chosen uniformly at random. Then, the first and last timestamps of the pattern's timeframe are sampled uniformly at random. Next, we select the streams that are to be included in the pattern. We use two alternative mechanisms for this task. DISTGEN emulates a realistic scenario for the creation of spatiotemporal patterns. It starts by randomly choosing the first stream $D_1$ to be included. Then, each additional streams is added with a probability proportional to its distance from $D_1$. On the other hand, RANDGEN randomly samples the number of streams to be included. It then samples as many streams uniformly at random.

Table 4: List of Major Events between September of 2008 and July of 2009, from www.wikipedia.com. The 2nd column contains the query chosen for the event. The 3rd column shows a brief description of the event.

| # | Query | Event Description |
|---|---|---|
| 1 | Obama | Events regarding the actions of B. Obama, the new President of the USA since January of 2009. |
| 2 | financial crisis | Events regarding the global financial crisis. |
| 3 | terrorists | Events regarding terrorism. |
| 4 | Jackson | American entertainer Michael Jackson passes away. |
| 5 | swine | Events regarding the 2009 swine flu pandemic. |
| 6 | earthquake | Events regarding earthquakes. |
| 7 | gaza | Events regarding the Israeli Palestinian conflict in the Gaza Strip. |
| 8 | ceasefire | Israel announces a unilateral ceasefire in the Gaza War. |
| 9 | yemenia | Yemenia Flight 626 crashes off the coast of Moroni, Comoros, killing all but one of the 153 passengers and crew. |
| 10 | piracy | Events regarding incidents of Piracy off the Somali coast. |
| 11 | Air France | Air France Flight 447 from Rio de Janeiro to Paris crashes into the Atlantic Ocean killing all 228 on board. |
| 12 | bush fires | Deadly bush fires in Australia kill 173, injure 500 more, and leave 7,500 homeless. |
| 13 | Nkunda | Congolese rebel leader L. Nkunda is captured by Rwandan forces. |
| 14 | Vieira | The President of Guinea-Bissau, J. B. Vieira, is assassinated. |
| 15 | Tsvangirai | M. Tsvangirai is sworn in as the new Prime Minister of Zimbabwe. |
| 16 | Rajoelina | Andry Rajoelina becomes the new President of Madagascar after a military coup d'etat. |
| 17 | Fujimori | Former Peruvian Pres. Fujimori is sentenced to 25 years in prison for killings and kidnappings by security forces. |
| 18 | Zelaya | The Supreme Court of Honduras orders the arrest and exile of President M. Zelaya. |

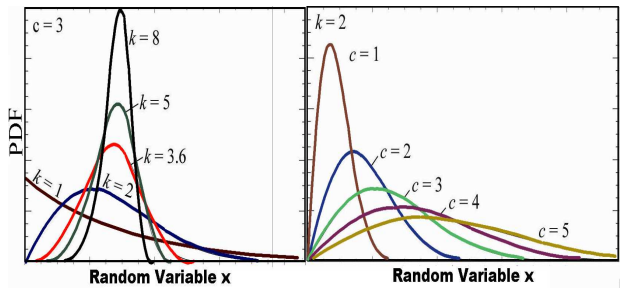

**Figure 9:** PDF curves for the weibull distribution, for different values of the shape (k) and scale (c) parameters.

Finally, we need to inject the extra frequency that is due to the created patterns into the streams. Each pattern captures the burstiness of a term relevant to an event. Following the event's occurrence, the term's burstiness builds up until it reaches a peek. It then starts to deflate until finally the impact of the event wears off. Depending on the nature of the event, burstiness can rise sharply (e.g. an unexpected earthquake), or build up slowly (e.g. the occurrence of big sports game) to reach its peek. Similarly, it may persist for an extended timeframe or quickly dissipate.

Taking the above into consideration, we utilize the Weibull Distribution to sample frequency values. The density function of this distribution emulates the burstiness process, and is thus ideal for our purpose. By setting the distribution's *shape* (k) and *scale* (c) parameters, we can tune the curve to emulate virtually any type of frequency pattern. The PDF for a random variable $x$ is computed as follows:

$$f(x;c,k) = \begin{cases} \frac{k}{c}\left(\frac{x}{c}\right)^{k-1} e^{-(x/c)^k} & x \geq 0 \\ 0 & x < 0 \end{cases} \quad (12)$$

Figure 9 shows examples of PDF curves for different values of these parameters. As can be seen in the Figure, the curve can be tuned to emulate the progress of virtually every type of event. This includes, for example, events that slowly build up and fade away or unexpected bursty events. In our case, the random variable $x$ is the order of the timestamp (i.e. $1, 2, ..., |T|$). The respective PDF values are then the desired frequency values. The highest point of the curve is the distribution's mode $m$ (i.e. the most frequent value). Thus, we can easily set the frequency $P$ at which the curve peeks to any given value $v$, by simply multiplying all the values in the sequence with $v/m$.

In realistic scenarios, the frequency pattern of the same event may differ from stream to stream. To account for this in our experiments, the values for $c, k, P$ are chosen uniformly at random for each stream, to ensure high variability in the produced patterns.

## C. THE GETMAX ALGORITHM

GetMax reads the values from left to right. Segments that are candidates for maximality are kept in a list. For each candidate $I_j$, we record the sum $l_j$ of all scores up to the leftmost score of $l_j$ (exclusive) and the sum $r_j$ up to the rightmost score of $I_j$ (inclusive). Non-positive scores require no special handling. If a positive score is read, a new sequence $I_k$ containing only this score is created and processed as follows:

1. Search the list from right to left for the maximum value of $j$ satisfying $l_j < l_k$.

2. If there is no such $j$ or there is such a $j$ and $r_j \geq r_k$, append $I_k$ to the list. If there is such a $j$ but $r_j < r_k$, extend $I_k$ up to the leftmost score in $I_j$ (inclusive). Remove candidates $I_j, I_{j+1}, ..., I_{k-1}$ from the list and consider $I_k$ (now numbered $I_j$) from step 1.

After the entire input has been processed, the candidates left in the list are the maximal segments. We refer the reader to the original paper [21] for a more detailed analysis.